\documentclass[english,aps,prb,twocolumn,floats,showpacs]{revtex4}
\usepackage[T1]{fontenc}
\usepackage[latin9]{inputenc}
\usepackage{amsmath}
\usepackage{amssymb}
\usepackage{graphicx}
\usepackage{esint}

\makeatletter

\@ifundefined{textcolor}{}
{%
 \definecolor{BLACK}{gray}{0}
 \definecolor{WHITE}{gray}{1}
 \definecolor{RED}{rgb}{1,0,0}
 \definecolor{GREEN}{rgb}{0,1,0}
 \definecolor{BLUE}{rgb}{0,0,1}
 \definecolor{CYAN}{cmyk}{1,0,0,0}
 \definecolor{MAGENTA}{cmyk}{0,1,0,0}
 \definecolor{YELLOW}{cmyk}{0,0,1,0}
}

\usepackage{babel}
\usepackage{bm}

\makeatother

\begin{document}

\title{Quantum Collapse of a Magnetic Skyrmion}

\author{Amel Derras-Chouk, Eugene M. Chudnovsky, and Dmitry A. Garanin}

\affiliation{Physics Department, Herbert H. Lehman College and Graduate School, The City University of New York \\
 250 Bedford Park Boulevard West, Bronx, New York 10468-1589, USA}

\date{\today}
\begin{abstract}
Quantum collapse of a small skyrmion in a thin magnetic film with Dzyalishinskii-Moriya (DMI) interaction has been studied. The energy of the skyrmion and the stability threshold determined by the DMI, the external magnetic field, and the underlying atomic lattice are investigated analytically and numerically. The Lagrangian describing the coupled dynamics of the skyrmion size and the chirality angle is derived. Equations of motion possess an instanton solution that corresponds to the skyrmion underbarrier contraction via quantum tunneling with subsequent collapse and decay of the topological charge. The tunneling rate is computed and the conditions needed to observe quantum collapse of a skyrmion in a magnetic film are discussed. 
\end{abstract}


\maketitle

\section{Introduction}
Skyrmions have been originally introduced in nuclear physics as possible nonlinear field-theory prototypes of hadrons \cite{SkyrmePRC58}. They quickly attracted interest in condensed matter physics in applications to topological defects in 2D ferro- and antiferromagnetic films and layered materials \cite{BelPolJETP75,WiegmannPRL88,WenZeePRL88,HaldanePRL88,ChaHalNelPRB89,Lectures,Brown-book}, Bose-Einstein condensates \cite{AlkStoNat01}, quantum Hall effect \cite{SonKarKivPRB93,StonePRB93}, anomalous Hall effect \cite{YeKimPRL99}, and liquid crystals \cite{WriMerRMR89}. Skyrmions are currently at the forefront of research in magnetism due to their interesting topological properties and their potential for topologically protected information storage, see, e.g., reviews, Refs.\ \onlinecite{Nagaosa2013,Klaui2016,Leonov-NJP2016,Hoffmann-PhysRep2017,Fert-Nature2017}. 

Solid-state research on skyrmions is focusing on their stability, dynamics and various symmetry properties. The majority of published works treated skyrmions as classical objects, although some attention has been paid to their quantum excitations as well \cite{Roldan-PRB2015}. In most cases application of classical theory is justified because even the smallest nanometer-size skyrmion would be comprised of a macroscopic number of degrees of freedom. Consequently, if one is interested in the dynamics of the skyrmion as a whole and not in small excitations, the field-theory action associated with such dynamics would be large compared to the Planck constant. This makes quantum features strongly suppressed. Nevertheless, they can be important in the context of information storage if the topological charge of a skyrmion can decay in the long run via quantum processes. 

Besides its practical importance, if such a behavior of a skyrmion was detected in experiment, it would manifest another fascinating example of macroscopic quantum tunneling (MQT) that has been intensively studied in condensed matter physics in the past. MQT research included tunneling of the magnetic moment in single-domain magnetic nanoparticles and tunneling of domain walls \cite{MQT-book}, spin tunneling in molecular magnets \cite{Springer,Wernsdorfer}, tunneling of vortices in 2D superconductors and quantum depinning of flux lines in 3D superconductors \cite{Blatter}, tunneling between supercurrent states in nano-SQUIDS \cite{Clarke}, etc. 

In this paper we are asking a question whether quantum collapse of a classically stable skyrmion can be observed in experiment. In a pure exchange model in a 2D crystal, skyrmions collapse classically due to the violation of the scale invariance by the presence of the discrete atomic lattice \cite{CCG-PRB2012}. Anisotropy, dipole-dipole interaction (DDI), magnetic field, and confined geomery can stabilize significantly large magnetic bubbles with skyrmion topology \cite{IvanovPRB06,IvanovPRB09,Moutafis-PRB2009,Ezawa-PRL2010,Makhfudz-PRL2012}, while stability of small skyrmions requires other than Heisenberg exchange coupling, strong random anisotropy, or a non-centrosymmetric system with large Dzyaloshinskii-Moriya interaction (DMI) \cite{AbanovPRB98,Bogdanov-Nature2006,Heinze-Nature2011,Leonov-NatCom2015,Lin-PRB2016,Leonov-NJP2016,EC-DG-NJP2018}.

In this paper we study skyrmions stabilized by the DMI and an external magnetic field, which is a typical situation in most experiments. At a given strength of the DMI the size of the skyrmion is determined by the magnetic field; the stronger the field the smaller the size. Stable skyrmions above critical size $\lambda_c$ exist below a critical field $H_c$ determined by the strength of the DMI. Above that field skyrmions collapse irreversibly. We assume that skyrmion stability is dominated by the exchange, DMI, and strong magnetic field, and neglect the effect of the weaker dipolar fields that would make quantum problem significantly more involved. Classical dynamics of a collapsing skyrmion has been studied in Ref.\ \onlinecite{CCG-PRB2012}. For a nanometer size skyrmion it occurs on a nanosecond time scale. It has been shown that the skyrmion preserves its topological charge until the final stage of the collapse where it reaches an atomic size. At that point the topological charge of the skyrmion abruptly changes from 1 to 0. 

The existence of the critical field in a system with DMI allows one to control the energy barrier for the collapse of a stable skyrmion by tuning the magnetic field close to $H_c$. We show that such tuning of the field, which is easily within experimental reach, allows one to achieve a sufficiently large rate of underbarrier quantum contraction of the skyrmion below the critical size. Once such a process occurs due to quantum tunneling the skyrmion continues to collapse classically until it decays completely into the magnons  \cite{CCG-PRB2012}. We solve the tunneling problem by reducing it to the dynamics of a single parameter -- the skyrmion size $\lambda$. We show that equations of motion for $\lambda$ in imaginary time possess an instanton solution that corresponds to the quantum tunneling from $\lambda > \lambda_c$ to $\lambda < \lambda_c$. The Euclidean action along the instanton trajectory gives the WKB exponent for the tunneling, while the attempt frequency related to small oscillations of $\lambda$ near the energy minimum provides the pre-exponential factor for the tunneling rate. 

This paper is organized as follows. The dependence of the various components of the skyrmion energy on its size is discussed in Section \ref{Energy}. The vicinity of the critical field is studied in Section \ref{Vicinity}. The Lagrangian and equations of motion are derived in Section \ref{lagrangian}. The rate of quantum collapse is obtained in Section \ref{Quantum}. The results and suggestions for experiments are discussed in Section \ref{Discussion}. 

\section{Energy}\label{Energy}
We consider the following Hamiltonian of a 2D spin system
\begin{eqnarray}
{\cal{H}} & = & -\frac{J}{2}\sum_{<ij>} {\bf S}_i \cdot {\bf S}_j  \nonumber \\
& + & A \sum_i [({\bf S}_i \times {\bf S}_{i + \hat{x}})\cdot \hat{x} + ({\bf S}_i \times {\bf S}_{i + \hat{y}})\cdot \hat{y}]. \nonumber \\
& - & H\sum_i S_{iz}
\end{eqnarray}
Here the first term represents the exchange interaction between the nearest neighbors, with ${\bf S}_i$ being the spin at the $i$-th site of the crystal lattice and $J$ being the exchange constant. The second term describes the Bloch-type Dzyaloshinskii-Moriya interaction (DMI) of strength $A$ in a non-centrosymmetric crystal \cite{Leonov-NJP2016}. For the N\'{e}el-type DMI it should be replaced with $A \sum_i [({\bf S}_i \times {\bf S}_{i + \hat{x}})\cdot \hat{y} - ({\bf S}_i \times {\bf S}_{i + \hat{y}})\cdot \hat{x}]$. The third term is the Zeeman interaction between the spins and the magnetic field applied in the $z$-direction, perpendicular to the $xy$-plane of the film. 

The field theory counterpart of the above Hamiltonian that one can obtain by switching from summation to integration according to $\sum_i \rightarrow \int dxdy/a^2$ (where $a$ is the lattice constant) is
\begin{eqnarray}\label{E-continuous}
{\cal{H}} & = & -\frac{1}{2} Ja^4 \int dxdy \left[ \left(\frac{\partial \tilde{\bf S}}{\partial x}\right)^2 + \left(\frac{\partial \tilde{\bf S}}{\partial y}\right)^2\right]  \nonumber \\ 
&+ & Aa^3\int dxdy \left[\left(\tilde{\bf S} \times \frac{\partial \tilde{\bf S}}{\partial x}\right)\cdot \hat{x} + \left(\tilde{\bf S} \times \frac{\partial \tilde{\bf S}}{\partial y}\right)\cdot \hat{y}\right], \nonumber \\
& - & H \int dxdy \, \tilde{S}_z
\end{eqnarray}
where $\tilde{\bf S}(x,y)$ is the spin field of constant density $S/a^2$. 

\begin{figure}
\vspace{0.5cm}
\includegraphics[width=75mm]{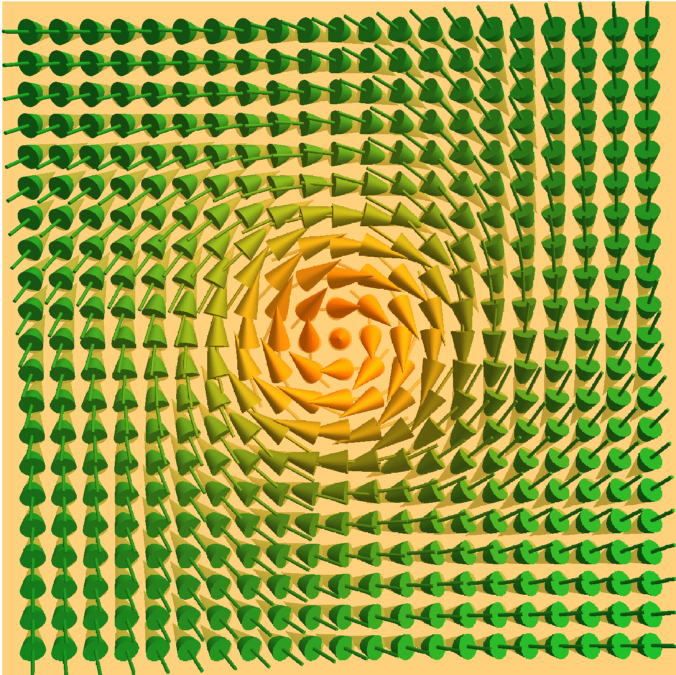}
\includegraphics[width=75mm]{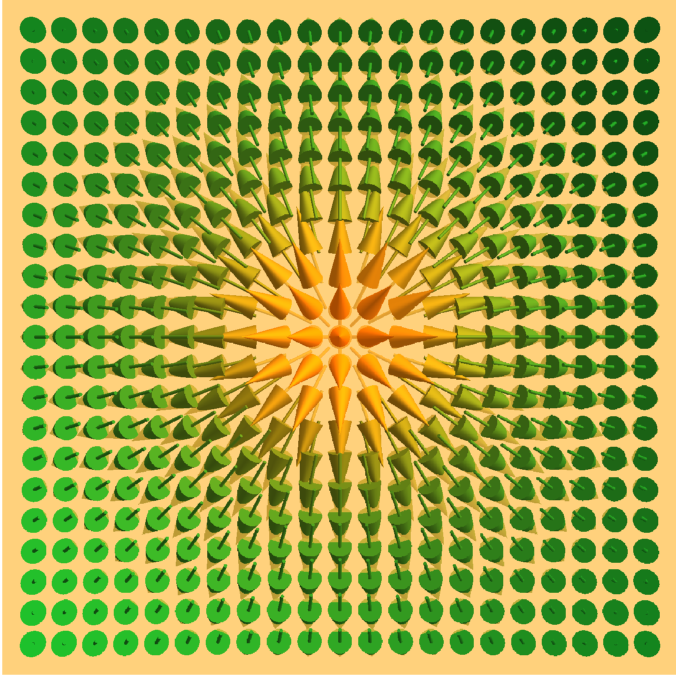}
\caption{Spin-field of the Bloch-type (upper panel) and the N\'{e}el-type (lower panel) BP skyrmions with $Q = 1$.}
\label{BP-skyrmions}
\end{figure}
Non-uniform configurations of the spin field in 2D are characterized by the topological charge, 
\begin{equation}\label{Q}
Q = \int \frac{d^2 r}{8\pi} \epsilon_{\alpha\beta} s_a \epsilon_{abc} \frac{\partial s_b}{\partial r_\alpha}\frac{\partial s_c}{\partial r_\beta} = \int \frac{dx dy}{4\pi} \: {\bf s}\cdot \frac{\partial {\bf s}}{\partial x} \times\frac{\partial {\bf s}}{\partial y},
\end{equation}
that takes integer values $Q = 0, \pm 1, \pm 2 , ...$. Here ${\bf s}$ is a unit vector specifying the direction of the spin field, ${\bf s} = \tilde{\bf S}/\tilde{S}$. At $H = 0$ and $A = 0$ the non-uniform rotations of the spin field that minimize the exchange energy are Belavin-Polyakov (BP) skyrmions \cite{BelPolJETP75}. For example, for $Q = 1$ the components of ${\bf s}$ are given by 
 \begin{equation}\label{BP}
s_x  =  2\lambda \frac{r\cos(\phi + \gamma)}{r^2 + \lambda^2}, \;
s_y  =  2\lambda \frac{r\sin(\phi + \gamma)}{r^2 + \lambda^2}, \;
s_z  =  \frac{\lambda^2 - r^2}{\lambda^2 + r^2},
\end{equation}
where ${\bf r} = (r \cos\phi, r\sin\phi)$  is the radius-vector in the $xy$-plane, $\lambda$ can be interpreted as the skyrmion size, and $\gamma$ is the chirality angle. In a continuous spin-field approximation the scale invariant 2D exchange energy of a $Q = 1$ BP skyrmion, $E_{ex} = 4\pi JS^2$, is independent of its size, which is confirmed by substitution of Eqs.\ (\ref{BP}) in the first term of Eq.\ (\ref{E-continuous}). The spin-field in the $Q = 1$ N\'{e}el-type ($\gamma = 0$) and Bloch-type ($\gamma = \pi/2$) skyrmions is shown in Fig.\ \ref{BP-skyrmions}. 

In practice one typically has $H \ll AS \ll JS$, so that the energy of the short-range rotations of $\tilde{\bf S}$ is dominated by the exchange. This suggests that for sufficiently small $\lambda$ and not very large $r$ Eqs.\ ({\ref{BP}) provide a good approximation for the skyrmion shape. Indeed, the dimensional analysis of the energies in Eq.\ (\ref{E-continuous}) shows that the DM energy of the BP skyrmion scales as $AS^2 (\lambda/a)$, while its Zeeman energy (up to a logarithm) scales as $HS(\lambda/a)^2$. Both are small compared to the exchange energy of the BP skyrmion when ${\lambda}/{a} \ll {J}/{A}$ and ${\lambda} \ll \delta_H$, with $\delta_H \equiv \sqrt{{JS}/{H}}a$. When these conditions are satisfied, interactions other than the exchange may affect skyrmion shape at $r \gg \lambda$. However at distances $r \lesssim \lambda$, the skyrmion shape is determined by the exchange interaction and is close to the BP shape given by Eqs.\ (\ref{BP}). This is confirmed by our numerical studies of skyrmions on spin lattices, see Appendix. 

Violation of the scale invariance by a discrete atomic lattice leads to the $-(2\pi J S^2/3)(a/\lambda)^2$ contribution to the energy of the BP skyrmion that forces it to collapse with a lifetime proportional to $(\lambda/a)^5$ in the absence of any other stabilizing interactions \cite{CCG-PRB2012}. The DMI and Zeeman interaction can stabilize skyrmions. Substitution of Eqs.\ (\ref{BP}) into Eq.\ (\ref{E-continuous}) with the addition of the discrete-lattice contribution gives the following dependence of the skyrmion energy on size and chirality angle: 
\begin{equation}\label{E}
E = -\frac{2\pi J S^2}{3\bar{\lambda}^2}  - 4\pi A S^2 \bar{\lambda} \sin\gamma + 4\pi HS \bar{\lambda}^2 l(H,\bar{\lambda}),
\end{equation}
where we have dropped the dominant exchange contribution, $4\pi JS^2$, that does not depend on $\lambda$ and introduced dimensionless $\bar{\lambda} = \lambda/a$. The function $l(H,\bar{\lambda})$, having logarithmic dependence on $\delta_H/\lambda$, is given by Eq.\ (\ref{num-Zeeman}) of the Appendix. For certainty we consider a Bloch-type DMI with $A > 0$, but all formulas can be modified in a trivial manner for other types of DMI. The plus sign of the last (Zeeman) term in Eq.\ (\ref{E}) comes from the direction of the field being opposite to the magnetic moment of the skyrmion. The first (lattice) term and the last (Zeeman) term in the energy favor collapse of the skyrmion, while the second (DMI) term favors expansion of the skyrmion and $\gamma = \pi/2$. This provides the energy minimum on $\lambda$ at $H < H_c$. The dependence of the skyrmion energy on the skyrmion size, $\lambda$, for fields close to $H_c$ is shown in Fig.\ \ref{E-lambda}. 

\begin{figure}
\includegraphics[width=87mm]{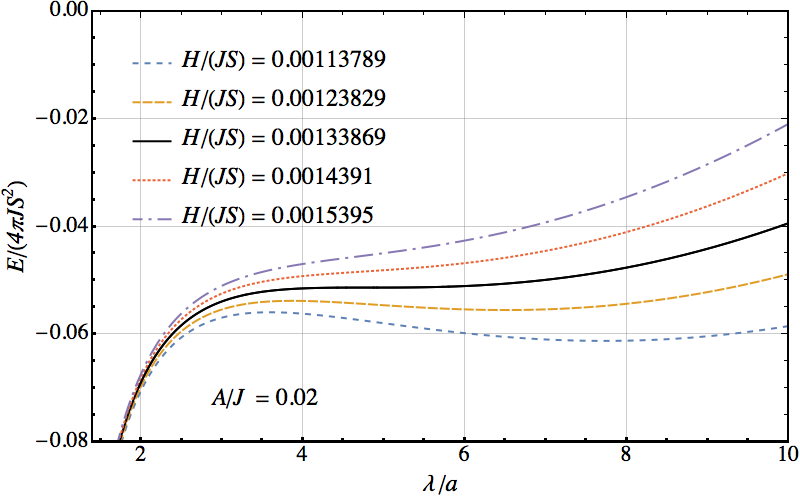}
\caption{Dependence of skyrmion energy on skyrmion size given by Eq.\ (\ref{E}) near the critical (collapse) field.}
\label{E-lambda}
\end{figure}
To increase confidence in our results we will develop and compare three approaches. At first we will treat the function $l$ as a constant. This is justified when $dl/d\lambda$ contributes little to the derivative with respect to $\lambda$ of the Zeeman energy, $4\pi HS \bar{\lambda}^2 l(H,\bar{\lambda})$, which requires $\delta_H \gg \lambda$, $2l \gg 1$. Such approximation elucidates the key features of the tunneling problem and it provides the leading dependence of the tunneling rate on parameters. It will be further improved by the numerical solution of the continuous problem that takes into account logarithmic corrections and also by computation of all energies in a discrete model on 500$\times$500 spin lattices. 

\begin{figure}
\includegraphics[width=8.7cm,angle=0]{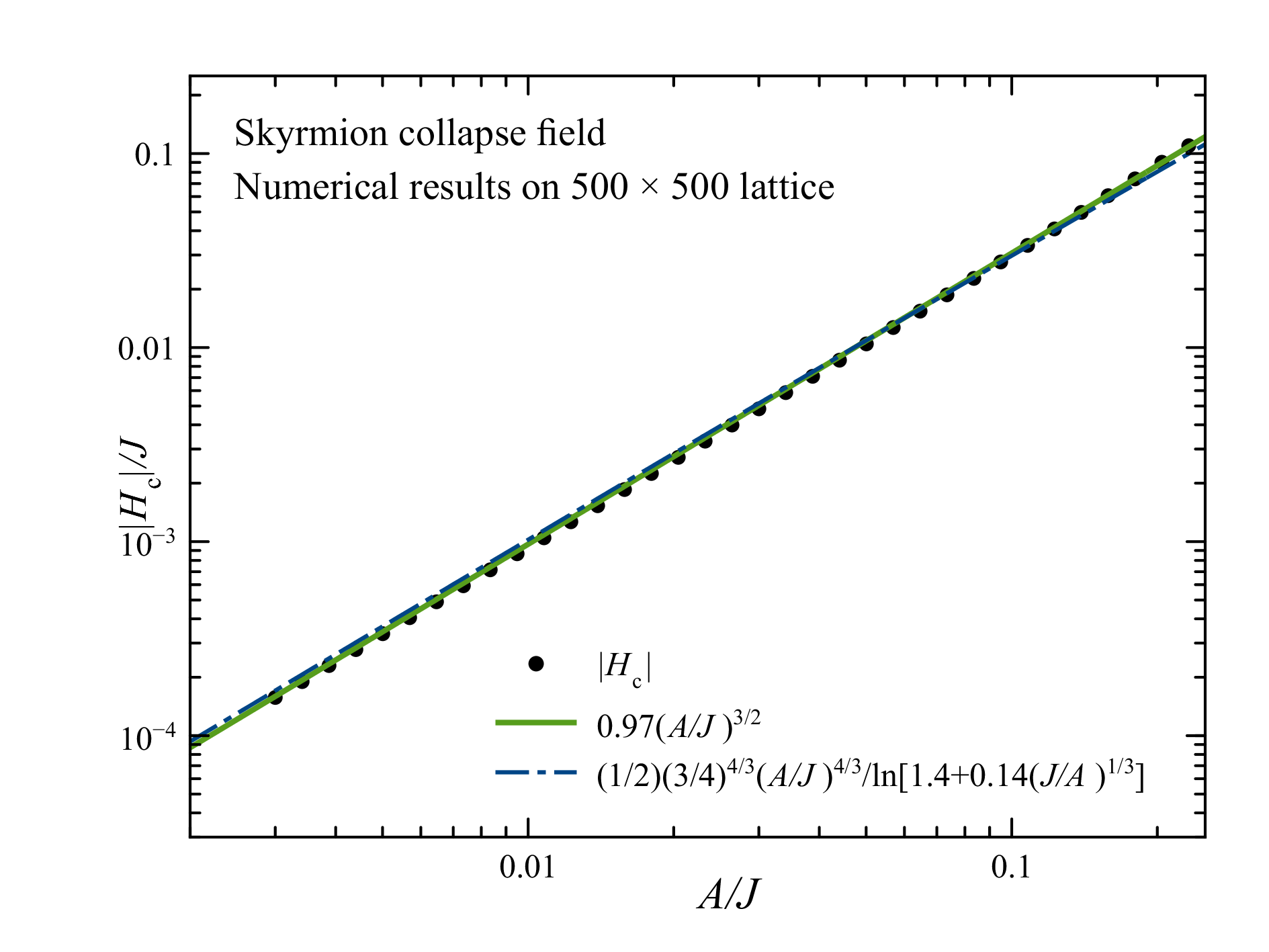}
\includegraphics[width=8.7cm,angle=0]{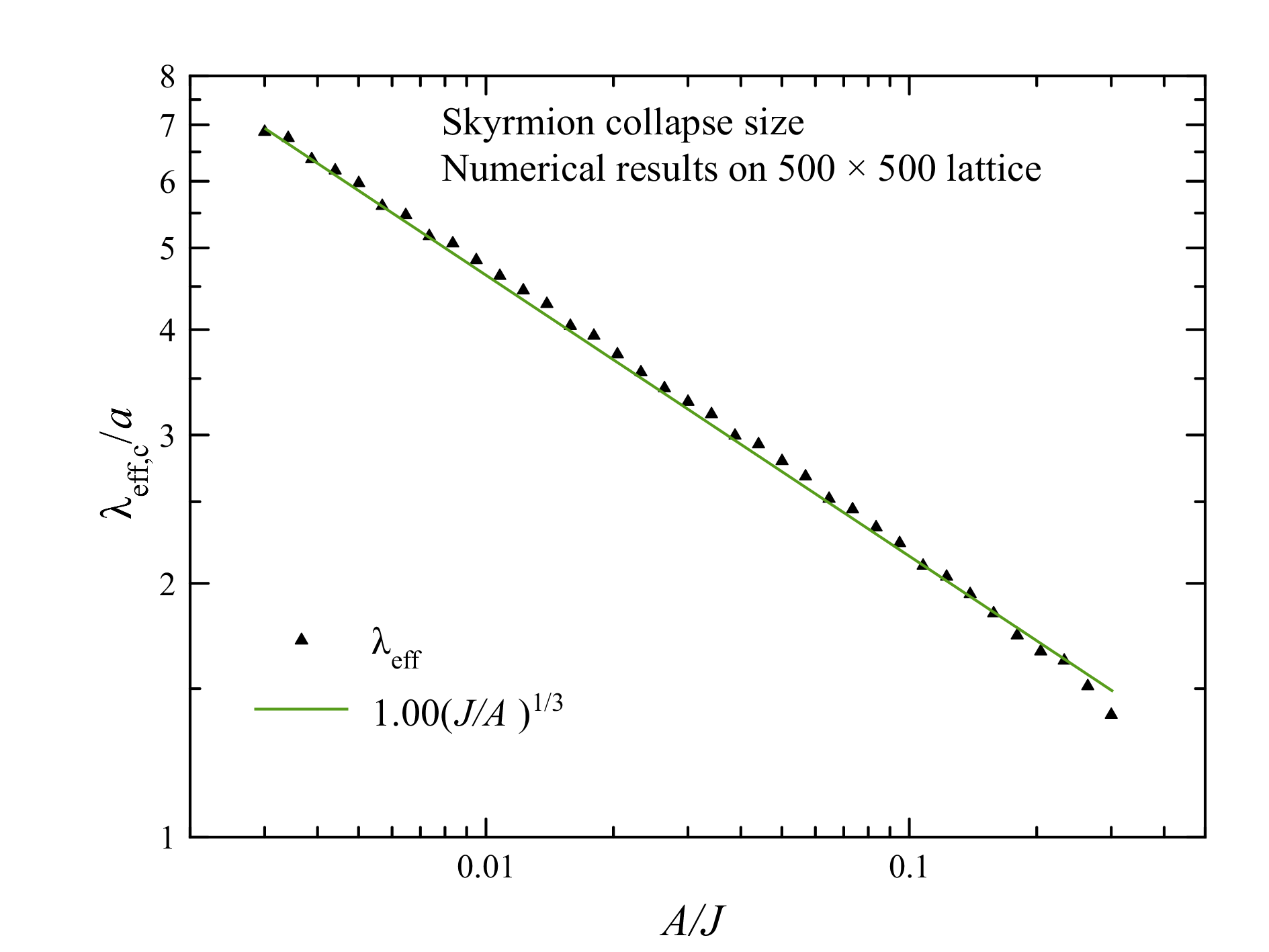}
\caption{Dependence of $H_c$ (upper panel) and $\lambda_c$ (lower panel) on $A/J$ obtained on 500 $\times$ 500 spin lattices, see Appendix for details.}
\label{scaling}
\end{figure}
If one neglects  logarithmic  dependence of $l$  on $\lambda$ the condition that the first and the second derivative of $E(\lambda)$ equal zero give the following dependence of the critical field $H_c$ and the critical skyrmion size $\lambda_c$ (at $H = H_c$) on parameters:
\begin{equation}\label{collapse}
\frac{H_c}{JS} = \frac{2^{1/3}}{16l\kappa^{4/3}}, \quad \bar{\lambda}_c  = (4\kappa)^{1/3}, \quad \kappa = \frac{J}{3A} \gg 1 .
\end{equation}
This approximation can be improved by assuming logarithmic dependence of $l$ on ${\kappa}$ in the expression for $H_c$. The scaling of $H_c/(JS)$ and $\lambda_c/a$ on $\kappa$ provided by Eqs.\ (\ref{collapse}) has been tested by finding these parameters numerically from Eq.\ (\ref{E}) without assuming $l = {\rm const}$, and also by computing the inflection point of the total energy numerically on spin lattices of large size, see Appendix for details. All three methods produce close results, see Fig.\ \ref{scaling} that represents the lattice method believed to be the most accurate one.  Note that in the numerical work $\kappa^{4/3}$ multiplied by a function that contains logarithmic dependence on $\kappa$ can easily be interpreted as a power of $\kappa$ that is slightly different from $4/3$, e.g., $3/2 =4/3 + 1/6$, and this is what is seen in the upper panel of Fig.\ \ref{scaling}.

\section{Vicinity of the critical field}\label{Vicinity}
It is expected that only the smallest skyrmions will have an appreciable rate of quantum collapse and only when the corresponding energy barrier is sufficiently small. Thus the quantum problem we are interested in must be studied near $ H = H_c$. In this region the energy shown in Fig.\ \ref{energy-critical} is given by
\begin{equation}\label{E-delta}
\frac{E}{JS^2} = \frac{\pi}{2^{1/3} 3  \kappa^{5/3}}  \left[|\bar{\delta}_t|(\delta\bar{\lambda})^2 + (\delta\bar{\lambda})^3 \right],
\end{equation}
where  $\delta \bar{\lambda} = \bar{\lambda} - \bar{\lambda}_1$, $\bar{\delta}_t = \bar{\lambda}_t - \bar{\lambda}_1 = - 2^{1/6} 4 \kappa^{1/3}\epsilon^{1/2}$, and 
\begin{equation}
\epsilon = 1 - \frac{H}{H_c} \ll 1.
\end{equation}
The energy minimum corresponds to $\gamma = \pi/2$, any departure of $\gamma$ from $\pi/2$ is related to the time derivatives of $\lambda$. In that sense Eq.\ (\ref{E-delta}) gives the potential energy of the skyrmion in the field just below $H = H_c$, shown in Fig.\ \ref{energy-critical}. 
\begin{figure}
\vspace{-0.55cm}
\includegraphics[width=8.7cm,angle=0]{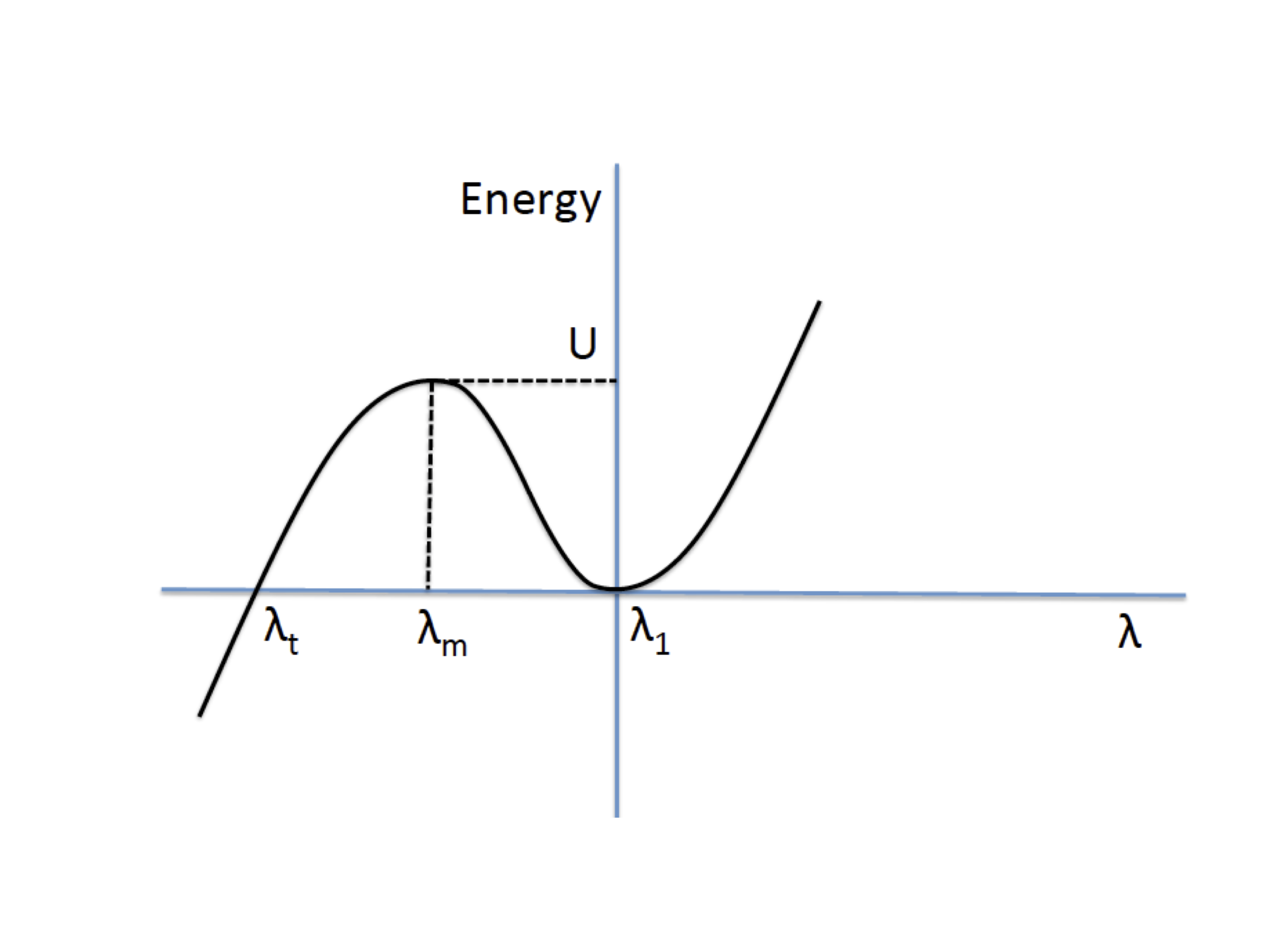}
\vspace{-1cm}
\caption{Schematic representation of the energy of the skyrmion below the critical field that shows the energy barrier $U$ and parameters $\lambda_1$, $\lambda_m$, $\lambda_t$ used in the text.}
\label{energy-critical}
\end{figure}

At any $H < H_c$ the reduced skyrmion size $\bar{\lambda}_1$ corresponding to the energy minimum satisfies
\begin{equation}\label{lambda1}
-1 +\frac{\kappa}{\bar{\lambda}_1^3} + \bar{\omega}_0 \bar{\lambda}_1 = 0,
\end{equation}
where  $\bar{\omega}_0 = {2 lH}/(AS)$. Close to the critical field 
\begin{equation}
\bar{\omega}_0 = \bar{\omega}_c (1 - \epsilon), \quad \bar{\omega}_c = \frac{3}{4\bar{\lambda_c}} = \frac{3}{4(4\kappa)^{1/3}}
\end{equation}
and the characteristic values of $\lambda$ shown in Fig.\ \ref{energy-critical} are 
\begin{eqnarray}
\bar{\lambda}_1 & = & \bar{\lambda}_c + 2^{1/6}\kappa^{1/3}\epsilon^{1/2} \\
\bar{\lambda}_m & = & \bar{\lambda}_c - 2^{1/6} 2 \kappa^{1/3}\epsilon^{1/2} \\
\bar{\lambda}_t & = & \bar{\lambda}_c - 2^{1/6} 3 \kappa^{1/3}\epsilon^{1/2}.
\end{eqnarray}
The energy barrier, $U = E(\lambda_m) - E(\lambda_1)$,  is given by 
\begin{equation}
\frac{U}{JS^2}  =  \frac{4\pi \, 2^{1/6} }{3\kappa^{2/3}}\,\epsilon^{3/2}.
\end{equation}
Note that close to $H_c$ these results, obtained under the assumption $l = {\rm const}$, do not depend on $l$ and, therefore, must be valid regardless of that assumption. The dependence of $l$ on $\lambda$ and $H$ simply renormalizes $H_c$, preserving the dependence of all variables on $\epsilon = 1 - H/H_c$ while yielding $l = l_c$ independent of $\epsilon$.

\section{Lagrangian and equations of motion}\label{lagrangian}
To solve the tunneling problem we need the dynamical equations for $\bar{\bf S}(r,t)$.  In the absence of dissipation they are given by
\begin{equation}\label{LL}
\hbar\frac{\partial \bar{\bf S}}{\partial t} = \bar{\bf S} \times {\bf B}_{\rm eff}, \quad {\bf B}_{\rm eff} = -\frac{\delta {\cal{H}}}{\delta \bar{\bf S}}.
\end{equation}
Writing components of the spin field in terms of the angles in spherical coordinates gives:
\begin{eqnarray}
\bar{S}_x & = & \frac{S}{a^2} \sin\Theta(x,y) \cos\Phi(x,y) \\
\bar{S}_y & = & \frac{S}{a^2} \sin\Theta(x,y) \sin\Phi(x,y) \\
\bar{S}_z & = & \frac{S}{a^2} \cos\Theta(x,y).
\end{eqnarray}
Noticing that $\hbar \bar{S}_z$ and $\Phi$ form a canonically conjugate pair of the generalized momentum and generalizede coordinate, it is easy to see that Eqs.\ (\ref{LL}) written in components of $\bar{\bf S}$ follow from the Lagrangian
\begin{equation}\label{Lagrangian}
{\cal{L}} = \hbar S \int \frac{dxdy}{a^2}  \dot{\Phi}(\cos\Theta + 1) - {\cal{H}}.
\end{equation}
Here $1$ is added to $\cos\Theta$ to make the first term zero at infinity where $\cos\Theta = -1$. This adds a total time derivative to the  Lagrangian that does not affect classical equations of motion. It may, however, contribute a phase, $S\Delta\Phi$, to the amplitude of quantum transition if $\Phi$ changes by $\Delta\Phi$ when going from the initial to the final state. Note that the first integrated quantity in Eq.\ (\ref{Lagrangian}) has a geometrical meaning  \cite{MQT-book} of the surface area swept by the spin field in a closed path around the south pole of the sphere of radius $\bar{S}$. 

According to Eqs.\ (\ref{BP}) for the skyrmion
\begin{eqnarray}
\tan\Phi & = & \frac{s_y}{s_x}  = \tan(\phi + \gamma), \quad \dot{\Phi} = \dot{\gamma} \\
\cos\Theta + 1&  = & s_z + 1 = \frac{2\lambda^2}{\lambda^2 + r^2}.
\end{eqnarray}
Substituting this into Eq.\ (\ref{Lagrangian}) and replacing ${\cal{H}}$ with $E$ of Eq.\ (\ref{E}) one obtains 
\begin{equation}
{\cal{L}} = 4\pi S(\hbar \dot{\gamma} - H)\bar{\lambda}^2  l(H,\bar{\lambda}) + \frac{2\pi J S^2}{3\bar{\lambda}^2} + 4\pi A S^2 \bar{\lambda} \sin\gamma .
\end{equation}
The equations of motion are
\begin{equation}\label{motion}
\frac{\partial {\cal{L}}}{\partial \bar{\lambda}} = 0, \quad \frac{d}{dt} \frac{\partial {\cal{L}}}{\partial \dot{\gamma}} = \frac{\partial {\cal{L}}}{\partial {\gamma}}.
\end{equation}

If one treats the logarithm as a constant $l$ the equations of motion become
\begin{eqnarray}\label{eq-motion}
&& \frac{d \bar{\lambda}}{d \tau} = \cos\gamma  \nonumber \\
&& \bar{\lambda} \frac{d \gamma}{d \tau} = -\sin\gamma +\frac{\kappa}{\bar{\lambda}^3} + \bar{\omega}_0 \bar{\lambda},
\end{eqnarray}
where we have introduced dimensionless time $\tau = [AS/(2\hbar l)] t$. The minimum of the energy corresponds to a stationary solution of the above equations with $\gamma = \gamma_1 = \pi/2$ and $\lambda = \bar{\lambda}_1$ satisfying Eq.\ (\ref{lambda1}).
Writing near the minimum 
\begin{equation}\label{deviations}
\gamma = \frac{\pi}{2} +\delta\gamma, \qquad \bar{\lambda} = \bar{\lambda}_1 + \delta \bar{\lambda}
\end{equation}
one obtains the following linearized equations
\begin{eqnarray}\label{osc}
&& \frac{d \delta \bar{\lambda}}{d \tau} = -\delta\gamma \label{gamma-delta} \\
&& \frac{d \delta\gamma}{d \tau} = \frac{1}{\bar{\lambda}_1}\left(4\bar{\omega}_0 - \frac{3}{\bar{\lambda}_1}\right) \delta \bar{\lambda}
\end{eqnarray}
that describe harmonic oscillations of $\delta \gamma$ and $\delta \bar{\lambda}$ at a frequency
\begin{equation}\label{frequency}
\bar{\omega}_1 = \sqrt{\frac{1}{\bar{\lambda}_1}\left(4\bar{\omega}_0 - \frac{3}{\bar{\lambda}_1}\right)}.
\end{equation}
According to Eq.\ (\ref{lambda1}) $\bar{\lambda}_1 \rightarrow 1/\bar{\omega}_0$ at $H \rightarrow 0$. In this limit $\bar{\omega}_1 \rightarrow \bar{\omega}_0$ and $\bar{\omega_1} \tau \rightarrow (H/\hbar)t$, making the skyrmion size oscillate in real time at the ESR  frequency. As the magnetic field increases the oscillation frequency becomes smaller than the ESR frequency $H/\hbar$. It first increases with the field, but then decreases and becomes zero at $H = H_c$, where the energy minimum disappears. 

Close to the critical field one has 
\begin{equation}
\bar{\omega}_1 =  \frac{3^{1/2}2^{1/12}\epsilon^{1/4}}{2\kappa^{1/3}}.
\end{equation}
Note again independence of the reduced frequency $\bar{\omega}_1$ of $l$. The real-time frequency of the small oscillations of skyrmion size near the energy minimum for $\epsilon \ll 1$ is
\begin{equation}\label{frequency-epsilon}
\omega_{\epsilon} = \frac{3^{1/2}2^{1/12}\epsilon^{1/4}}{2\kappa^{1/3}}\left(\frac{|A|S}{2\hbar l}\right).
\end{equation}

\section{Quantum collapse of a skyrmion}\label{Quantum}
At $H$ close to $H_c$, one has  $ \bar{\lambda} = \bar{\lambda}_1 + \delta \bar{\lambda}$, $\gamma = \frac{\pi}{2} +\delta\gamma$, with $\delta\gamma$ satisfying Eq.\ (\ref{gamma-delta}).  Switching to the imaginary time, $u = it$, $\bar{u} = i\tau$, one obtains from the equations of motion (\ref{eq-motion}) the following equation for $\delta\lambda$: 
\begin{equation}\label{eq-imtime1}
 \frac{d^2\delta\bar{\lambda}}{d\bar{u}^2} =  \frac{1}{8\kappa}\left[2|\bar{\delta}_t|(\delta\bar{\lambda}) + 3(\delta\bar{\lambda})^3\right] .
\end{equation}
It is easy to see that this equation corresponds to the condition of constant total energy (\ref{E}) for the motion in imaginary time, which describes quantum tunneling of $\lambda$ from $\lambda_1$ to $\lambda_t$, see Fig.\ \ref{energy-critical}. By subtracting $E(\lambda_1)$ this energy can be made zero:
\begin{eqnarray}
&& \frac{E}{JS^2} =  \frac{4 \pi}{3\kappa}\left[\frac{3\epsilon^{1/2}}{2^{1/6} 4 \kappa^{1/3}}\delta\bar{\lambda}^2 + \frac{2\kappa}{\bar{\lambda}_c^5}\delta\bar{\lambda}^3 + \frac{\bar{\lambda}_c}{2}\delta\gamma^2\right] \\
& & =  \frac{\pi}{2^{1/3} 3  \kappa^{5/3}}  \left[|\bar{\delta}_t|(\delta\bar{\lambda})^2 + (\delta\bar{\lambda})^3 - 4\kappa \left(\frac{d\delta\bar{\lambda}}{d\bar{u}}\right)^2\right] = 0. \nonumber \\
\end{eqnarray}

Choosing reduced variables,
\begin{equation}
\delta\bar{\lambda} = \frac{\delta\lambda}{|\delta_t|},  \quad \bar{u}' = \frac{|\delta_t|^{1/2}}{2 \kappa^{1/2}}\bar{u}.
\end{equation}
Eq.\ (\ref{eq-imtime1}) can be written as
\begin{equation}\label{eq-imtime2}
 \left(\frac{d\delta\bar{\lambda}}{d\bar{u}'}\right)^2 =  (\delta\bar{\lambda})^2 + (\delta\bar{\lambda})^3.
\end{equation}
The solution is 
\begin{equation}
\delta\bar{\lambda} = \frac{-1}{\cosh^2(\bar{u}'/2)} ,
\end{equation}
that is, 
\begin{equation}
\delta\lambda = \frac{\delta_t}{\cosh^2(\omega_i u)}, \quad \omega_i = \frac{\omega_{\epsilon}}{2} = \frac{3^{1/2}2^{1/12}\epsilon^{1/4}}{4\kappa^{1/3}}\left(\frac{|A|S}{2\hbar l}\right).  
\end{equation}
It corresponds to the instanton (bounce trajectory in the imaginary time) that goes from $\lambda_1$ at $u = -\infty$ to $\lambda_t$ at $u = 0$ and back to $\lambda_t$ at $u = +\infty$, see Fig.\ \ref{energy-critical}. Note that $d\lambda/du$ is zero at $u = \pm\infty$ and on approaching $\lambda_t$ at $u = 0$. According to the equations of motion, $\gamma = \pi/2$ at $u = \pm\infty$ and $u =0$, and it is complex along the bounce trajectory under the barrier.

The tunneling rate is
\begin{equation}\label{Gamma}
\Gamma = Ae^{B}
\end{equation}
where $A \approx \omega_{\epsilon}/(2\pi)$ is the attempt frequency given by Eq.\ (\ref{frequency-epsilon})  and $B$ is the action integrated over the bounce, $B = (i/\hbar)\int dt {\cal{L}}$, which gives the WKB exponent for the tunneling \cite{MQT-book,Lectures}. In the absence of dissipation the equations of motion conserve energy. Adding a constant,  $E(\pi/2,\lambda_1)$, to the Lagrangian to have $E = 0$ at the energy minimum and along the bounce trajectory, one has 
\begin{equation}
B = \frac{i}{\hbar}\int dt (4\pi \hbar S l \dot{\gamma}\lambda^2) .
\end{equation}
For the instanton at $\epsilon \ll 1$ one has $\gamma = \pi/2 + \delta\gamma$ with $\delta\gamma$ satisfying the first of Eqs.\ (\ref{osc}), so that  $d\delta\gamma/d{\bar{u}}  = -i(d^2 \delta \lambda/d\bar{u}^2)$. This gives
\begin{equation}\label{eq-B}
B = 4\pi S l  i\int_{-\infty}^{+\infty}  d\bar{u} \frac{d\gamma}{d\bar{u}}\lambda^2 = 4\pi S l  \int_{-\infty}^{+\infty} d\bar{u} \frac{d^2\lambda}{d\bar{u}^2}\lambda^2 .
\end{equation}
where for the moment we treated $l$ as a constant. Integrating by parts, and using the condition $d\lambda/du = 0$ at the boundaries of the integration, one obtains for the integral in Eq.\ (\ref{eq-B})
\begin{eqnarray}
&& \int_{-\infty}^{+\infty} d\bar{u} \frac{d^2\lambda}{d\bar{u}^2}\lambda^2 = -2\int_{-\infty}^{+\infty}d\bar{u} \left(\frac{d\lambda}{d\bar{u}}\right)^2 \lambda   \nonumber \\
&& =-4\lambda_c \int_{0}^{|{\delta}_t|}d|{\delta\lambda|} \left(\frac{d|\delta\lambda|}{d\bar{u}}\right)   \nonumber \\
&& =-4\lambda_c \int_{0}^{|{\delta}_t|}d|{\delta\lambda|}\frac{1}{(4\kappa)^{1/2}}\left[|\delta_t|(|\delta\lambda|)^2 - (|\delta\lambda|)^3\right]^{1/2}  \nonumber \\
&& =-\frac{2^{1/12} 3^{1/2} 48}{5} \kappa^{2/3}\epsilon^{5/4} .
\end{eqnarray}
This gives
\begin{equation}\label{B}
B =  -\frac{2^{1/12}192\pi}{3^{1/6}5}Sl\left(\frac{J}{A}\right)^{2/3}\epsilon^{5/4} \approx -106\, Sl \left(\frac{J}{A}\right)^{2/3}\epsilon^{5/4} .
\end{equation}

Independent check of the scaling of the WKB exponent with $J/A$ and $\epsilon$ comes from noticing that for smooth potentials $B$ scales as $-U/(\hbar \omega_m)$ where $\omega_m$ is a frequency of small oscillations of the skyrmion size at the bottom of the inverted potential \cite{MQT-book}. A simple calculation shows that in our problem $\omega_m = \omega_{\epsilon}$. This frequency plays an important role in the tunneling problem \cite{EC-PRA}; the crossover from thermal overbarrier collapse of the skyrmion to thermally assisted quantum tunneling occurs at a temperature $T_c = \hbar\omega_m$. Below that temperature the skyrmion tunnels under the barrier from the energy levels $E < U$ to which it is thermally activated. 

To further check the above analytical results obtained by approximating the log in the Zeeman energy by a constant, we also solved the problem without making such an assumption by considering the Lagrangian  ${\cal{L}} = 4\pi S\hbar \dot{\gamma}\bar{\lambda}^2  l(H,\bar{\lambda}) - E$ with 
\begin{equation}
E = 4\pi SH\bar{\lambda}^2  l(H,\bar{\lambda}) - \frac{2\pi J S^2}{3\bar{\lambda}^2} -4\pi A S^2 \bar{\lambda} \sin\gamma .
\end{equation}
and finding the WKB exponent numerically with $l(H,\bar{\lambda})$ given by Eq.\ (\ref{num-Zeeman}). 

The dynamics of the skyrmion near the collapse field, $\epsilon = 1 - H/H_c \ll 1$, corresponds to $\gamma = \pi/2 + \delta\gamma$ with $|\delta\gamma| \ll 1$. This gives $\sin\gamma = 1 - \delta\gamma^2/2$,
\begin{equation}
E = 4\pi SH\bar{\lambda}^2  l(H,\bar{\lambda}) - \frac{2\pi J S^2}{3\bar{\lambda}^2} - 4\pi A S^2 \bar{\lambda} + 2\pi A S^2 \bar{\lambda}\delta\gamma^2 .
\end{equation}
From the second of Eqs.\ (\ref{motion}) one obtains
\begin{equation}
\delta\gamma = -\frac{\hbar}{AS\bar{\lambda}}\frac{d}{dt}\left[\bar{\lambda}^2  l(H,\bar{\lambda})\right] .
\end{equation}
Substitution into the energy gives
\begin{equation}
E =  4\pi SH\bar{\lambda}^2  l(H,\bar{\lambda}) - \frac{2\pi J S^2}{3\bar{\lambda}^2} - 4\pi A S^2 \bar{\lambda} + \frac{2\pi \hbar^2}{A\bar{\lambda}} \left[\frac{d(\bar{\lambda}^2 l)}{dt}\right]^2 .
\end{equation}
where the last term can be interpreted as the kinetic energy of the skyrmion. 

Parameters $\lambda_1$,  $\lambda_m$, and $\lambda_t$ should now be determined numerically by finding the minimum $\lambda_1$, the maximum $\lambda_m$, and the tunneling point $\lambda_t$ of the stationary energy in Fig. 1 given by
\begin{equation}
E_0 =  4\pi SH\bar{\lambda}^2  l(H,\bar{\lambda}) - \frac{2\pi J S^2}{3\bar{\lambda}^2} - 4\pi A S^2 \bar{\lambda} .
\end{equation}
The critical parameters $H_c$ and $\lambda_c$ should also be determined numerically from the condition that first and second derivatives of $E_0$ are zero. For the tunneling problem one should consider $E_0$ at $H = H_c(1 - \epsilon)$, with small $\epsilon$.  

Along the instanton trajectory  $E = E(\lambda_1)$. This gives ($u = it$)
\begin{equation}
 \frac{d(\bar{\lambda}^2 l)}{du} = -\frac{1}{\hbar}\left(\frac{A\bar{\lambda}}{2\pi}\right)^{1/2}\sqrt{E_0(\lambda) - E_0(\lambda_1)}
 \end{equation}
with the minus sign determined by the fact that $\lambda^2$ decreases when going from $\lambda_1$ to $\lambda_t$. For the WKB exponent one has
\begin{eqnarray}\label{full-B}
&& B  =  4\pi S i\int_{-\infty}^{+\infty}  d{t} \frac{d\delta\gamma}{d{t}}\bar{\lambda}^2l 
 = -4\pi S i \int_{-\infty}^{+\infty} d{t} \delta\gamma \frac{d(\bar{\lambda}^2 l)}{dt} \nonumber \\
& & =  - 4\pi S i \int d(\bar{\lambda}^2 l)\delta\gamma  =  \frac{4\pi\hbar i}{A \bar{\lambda}_c}\int d(\bar{\lambda}^2 l)\frac{d(\bar{\lambda}^2 l)}{dt} \nonumber \\
 && =  2 \left(\frac{2\pi}{A\bar{\lambda}_c}\right)^{1/2}\int d(\bar{\lambda}^2 l)\sqrt{E_0(\bar{\lambda}) - E_0(\bar{\lambda}_1)} \nonumber \\
&& =  - 4\left(\frac{2\pi}{A\bar{\lambda}_c}\right)^{1/2}\int_{\bar{\lambda}_t}^{\bar{\lambda}_1} d\bar{\lambda} \left[\frac{d(\bar{\lambda}^2 l)}{d\bar{\lambda}}\right]\sqrt{E_0(\bar{\lambda}) - E_0(\bar{\lambda}_1)} \nonumber \\
&& =  - 4\left(\frac{2\pi}{A\bar{\lambda}_c}\right)^{1/2}\left[\frac{d(\bar{\lambda}^2 l)}{d\bar{\lambda}}\right]_{\bar{\lambda}=\lambda_c}\int_{\bar{\lambda}_t}^{\bar{\lambda}_1} d\bar{\lambda}\sqrt{E_0(\bar{\lambda}) - E_0(\bar{\lambda}_1)} .\nonumber \\
\end{eqnarray}
The last step is taken by keeping in mind that $\lambda$ is close to $\lambda_c$ at $\epsilon \ll 1$. 

The limits of integration in Eq.\ (\ref{full-B}), as well as the integral, must be computed numerically. However, the scaling of $B$ on $J/A$ and $\epsilon$ can be seen right away by noticing that the integral Eq.\ (\ref{full-B}) is of order $|\bar{\lambda_1} - \bar{\lambda}_t|\sqrt{U}$. This immediately gives
\begin{equation}\label{B-exact}
B \propto Sl_c(J/A)\left(\frac{J}{A}\right)^{2/3}\epsilon^{5/4} .
\end{equation}
with $l_c(J/A)$ being the value of the log at the critical field, thus confirming the result of Eq.\ (\ref{B}). Numerically obtained dependence of the WKB exponent on parameters is shown in Fig.\ \ref{WKB}. 
\begin{figure}
\vspace{-0.3cm}
\includegraphics[width=8.7cm,angle=0]{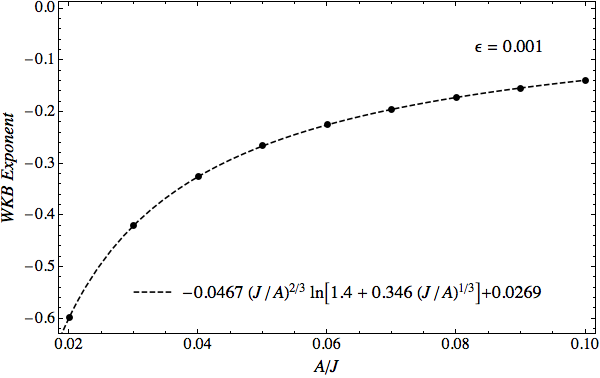}
\includegraphics[width=8.7cm,angle=0]{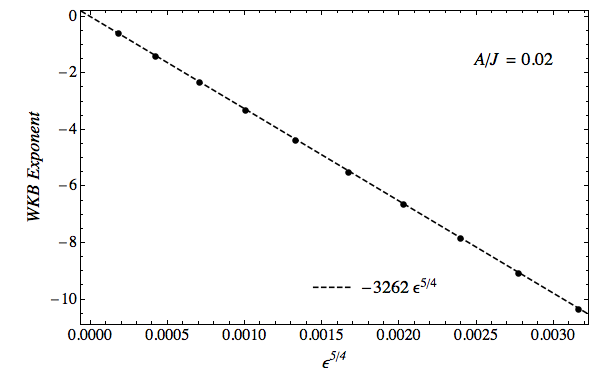}
\caption{Numerically computed dependence of the WKB exponent on $J/A$ (upper panel) and $\epsilon$ (lower panel). Its scaling with parameters confirms scaling obtained by other methods.}
\label{WKB}
\end{figure}
The agreement of all aproaches with each other provides good confidence in the results.  

\section{Discussion}\label{Discussion}

In the classical field theory skyrmions are topologically protected. Violation of  scale invariance by the atomic lattice makes small classical skyrmions collapse within a microscopic time. The combined effect of the DMI and the external magnetic field can stabilize skyrmions above a critical size $\lambda_c$. Quantum contraction of the skyrmion studied above conserves topological charge. Consequently, the skyrmion that emerges on the other side of the energy barrier with $\lambda_t < \lambda_c$ is the same BP skyrmion but of a smaller size and smaller magnetic moment. The change in the magnetic moment of the skyrmion that accompanies its quantum contraction is possible solely due to the DMI because it is the only interaction present in the problem that violates commutation of the Hamiltonian with $S_z$. This explains the presence of the power of $A$ in the denominator of the expression for the WKB exponent, Eqs.\ (\ref{B}) and (\ref{B-exact}). At $A \rightarrow 0$ one has $B \rightarrow -\infty$, resulting in the exponential smallness of $\Gamma$, that is, quantum tunneling of a skyrmion in our model is driven by the DMI. After the skyrmion contracts below $\lambda_c$ due to quantum tunneling it continues to contract towards $\lambda = 0$ in real time, radiating its energy into magnons \cite{CCG-PRB2012}. At the last stage of the collapse, when the skyrmion reaches the atomic size, it disappears with its topological charge jumping from $Q = 1$ to $Q = 0$. 

Not very close to $H_c$ the probability of quantum underbarrier contraction of the skyrmion, followed by its collapse, would generally be exponentially small owing to the large numerical factor in Eq.\ \ref{B}. It reflects the fact that even the smallest skyrmion would be rather macroscopic; that is, the action associated with it is large compared to $\hbar$. Indeed, the magnetic moment of a skyrmion of size $\lambda \approx \lambda_c$ would typically be in excess of $100\mu_B$. Quantum dynamics of systems comprised of a large number of degrees of freedom, such as, e.g., nano-SQUIDs and magnetic nanoparticles, have been observed in the past. Comparison with theory for such systems has been often hampered by the difficulty of preparing identical objects that exhibit tunneling. For example, quantum depinning of flux lines or 2D vortices in superconductors and of domain walls in ferromagnets depends on the local pinning potential that usually is widely distributed in magnitude. Studies of spin tunneling in single-domain magnetic particles always faced inability of experimentalists to prepare an array of identical nanoparticles. It received serious attention only after the discovery of resonant spin tunneling in magnetic molecules. Similar studies in superconductors are even more difficult as they require measurements of individual nano-SQUIDs. 

The advantage of small skyrmions for the studies of macroscopic quantum tunneling is that, similar to crystals of magnetic molecules, they can form an array of identical small magnetic objects. For, e.g., $J \sim 1000$K and $A/J \sim 0.02$ the critical collapse field $H_c$ should be around one tesla while the critical size of the skyrmion $\lambda_c$ should be on the order of four lattice spacings. As has been discussed above, the frequency of small oscillations of the skyrmion near the energy minimum should be generally in the ESR range and, therefore, easily observable. This frequency also determines the pre-exponential factor in the expression for the tunneling rate given by Eq.\ (\ref{Gamma}) as well as the crossover to thermally assisted quantum tunneling as the temperature is lowered. 

In this paper we neglected the dipole-dipole interaction (DDI) between the spins, as has been done for other tunneling problems in magnets. It is justified by the weakness of the DDI compared to all other interactions in the range of parameters used for the smallest skyrmions we are interested in. Incorporation of the DDI into our quantum tunneling model presents a challenge that we do not know how to address at this time. Another effect neglected by us is the effect of dissipation of the skyrmion motion on the probability of quantum collapse. In magnetic systems the effect of dissipation on tunneling of the magnetic moment is typically weak. It can be studied along the lines of Caldeira-Leggett aproach \cite{MQT-book}. The measure of the dissipation in magnetic materials is provided by the Landau-Lifshitz-Gilbert parameter which determines the FMR width and is usually small. For that reason, similar to the case of molecular magnets, it is unlikely that dissipation can significantly change our conclusions. 

To observe quantum tunneling on a time scale of a typical experiment the WKB exponent should be in the ballpark of 25-30. This can be achieved by applying the field close to $H_c$ as was done in experiments with individual magnetic particles in the past. 
According to Eq.\ (\ref{B}) for $A/J \sim 0.02$ it requires $\epsilon \sim 0.02$, that is, the field tuned within $100$G from the critical field, which is easily achievable. The onset of thermally assisted quantum tunneling must occur at $T_c = \hbar \omega_m$, which for the above parameters is in the ballpark of $1$K. The above numbers are given for a 2D monolayer of spins. Since the action is proportional to the number of layers, $N$, the WKB exponent in a multilayered film would change as $B \rightarrow NB$. Correspondingly, a smaller $\epsilon$, that is, the field closer to the collapse field will be required to observe quantum collapse of a skyrmion in thicker film. However, the temperature of the crossover from thermal overbarrier collapse to thermally assisted quantum collapse will remain the same.

We conclude with a notion that calculations presented here can be easily adjusted to other models and concrete materials chosen for experimental studies. Observation of the quantum collapse of a skyrmion, while challenging since it requires low temperatures and fine tuning of the field, appears to be within experimental reach and may even be less demanding than other MQT experiments performed to date. 

\section{Acknowledgements}

This work has been supported by the grant No. DE-FG02-93ER45487 funded by the U.S. Department of Energy, Office of Science.

\appendix

\section{Skyrmion energy on a spin lattice}\label{A1}

Numerical minimization of the skyrmion energy was performed on a $500\times500$
lattice using the method of Ref.\ \onlinecite{garchupro13} that consists
in successive rotations of spins at lattice sites $i$ in the direction
of the effective field $\mathbf{H}_{\mathrm{eff},i} = -\delta {\cal{H}}/\delta{\bf S}_i$ with the probability
$\alpha$ and \textit{overrelaxation} (i.e., flipping spins around
$\mathbf{H}_{\mathrm{eff},i}$) with the probability $1-\alpha$.
The first operaton reduces the energy of the system while the second
serves to better explore the phase space of the system via conservative
pseudo-dynamics, with $\alpha$ playing the role of the relaxation constant. The fastest energy minimization towards the deepest minum is achieved for $\alpha\ll1$. We use $\alpha=0.01$.

The skyrmion size $\lambda$ can be extracted from the numerical data
as \cite{CCG-PRB2012}
\begin{equation}
\lambda_{m}^{2}=\frac{m-1}{2^{m}\pi}a^{2}\sum_{i}\left(s_{iz}+1\right)^{m},
\end{equation}
where it is assumed that $s_{iz}=-1$ in the background and $s_{iz}=1$
at the center of the skyrmion. For the BP skyrmions
with $s_{z}$ given by Eq. (\ref{BP}), one has $\lambda_{m}=\lambda$
for any $m$. In this paper, we used $\lambda_{\mathrm{eff}}=\lambda_{4}$
to represent the numerically computed skyrmion size.

The numerical solution allows one to compute different contributions to
the equilibrium skyrmion energy, as well as $\lambda_{\mathrm{eff}}$,
for different values of the applied field $H$. Also, one can infer
a more general information via plotting the energy contributions in
the parametric form vs $\lambda_{\mathrm{eff}}$. The Zeeman energy
vs $\lambda_{\mathrm{eff}}$ is defined as $E_{Z}=-H\varDelta M_{\mathrm{skyrmion}}(\lambda_{\mathrm{eff}})$
with the skyrmion magnetic moment $\varDelta M_{\mathrm{skyrmion}}=\sum_{i}\left(s_{iz}+1\right),$
first computed numerically as a function of $H$ and then represented
parametrically via $\lambda_{\mathrm{eff}}$. The results are represented
in Fig. \ref{numerical}.
\begin{figure}
\includegraphics[width=8.7cm,angle=0]{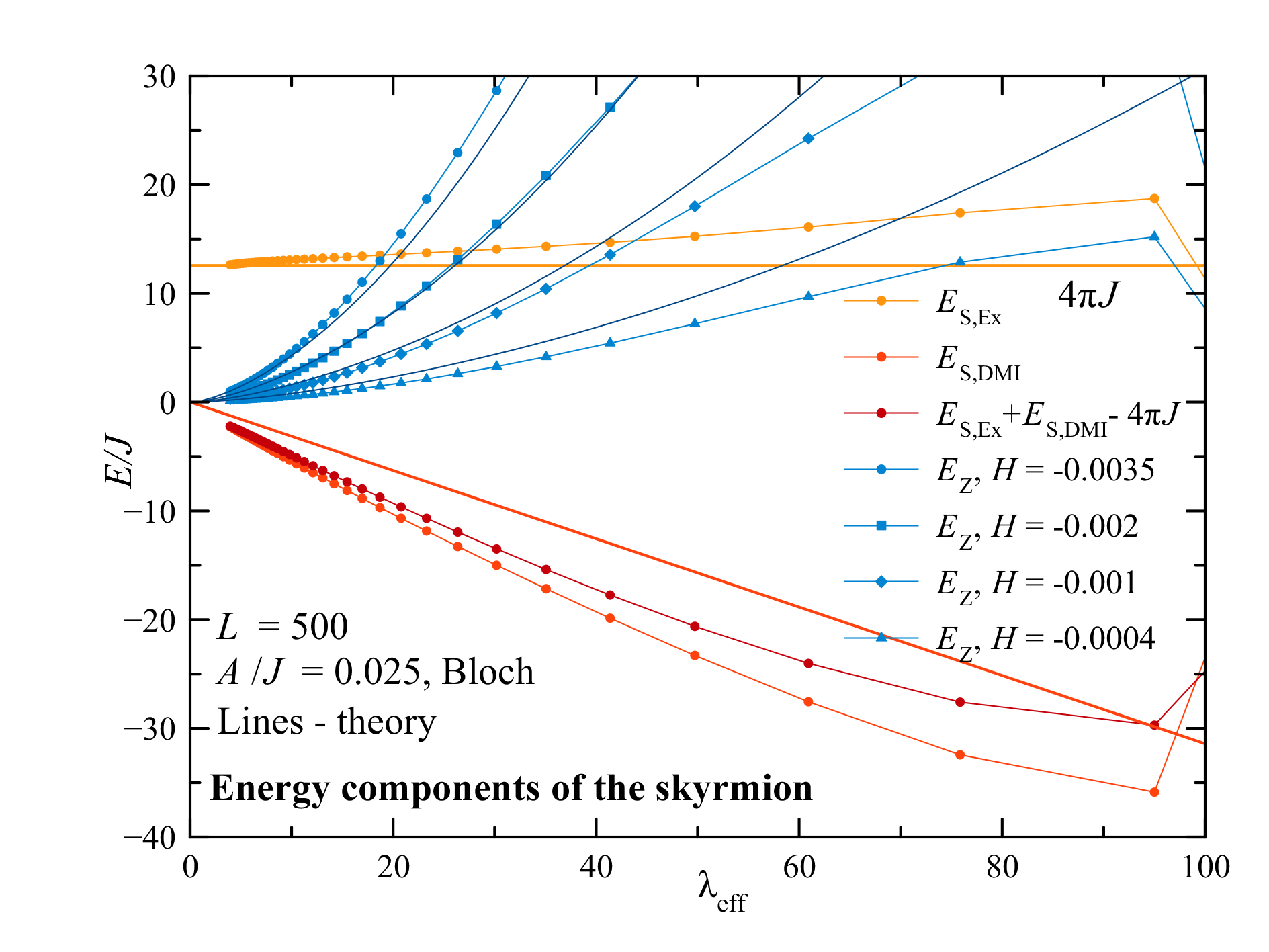}
\caption{Numerically computed exchange, DMI, and Zeeman energies of the $Q=1$
skyrmion on a $500\times500$ lattice with $S = 1$. Good agreement with analytical theory is seen, especially for smaller skyrmions that are of interest for the tunneling problem.}
\label{numerical} 
\end{figure}

One can see that for $\lambda_{\mathrm{eff}}$ below ten lattice spacings
the exchange energy of the skyrmion is close to the ground-state energy, $4\pi JS^2$, of the BP skyrmion. This is an
indication that such small skyrmions are close to the BP shape on
the scale $r\lesssim\lambda$ that dominates the exchange and the
DMI energies. The latter is due to the fact that these energies contain spatial
derivatives of the spin field. Any deformation of the BP shape on
the scale $r\lesssim\lambda$ would make the exchange energy higher.
According to Fig.\ \ref{numerical}, as $\lambda_{\mathrm{eff}}$
increases, the exchange energy slowly departs from $4\pi J$, indicating
more significant deformations of the skyrmion shape. Since the quantum
collapse problem is relevant for skyrmions of size well below $10a$
the BP shape must be a good approximation for the exchange and DMI
energies of such skyrmions.

For the Zeeman energy the situation is somewhat different. Since it
does not contain the derivatives of the spin field it is sensitive
to the shape of the skyrmion tail at $r\gg\lambda$ for which Zeeman
interaction becomes the dominant one. The scaling of the numerically
computed Zeeman energy, see Fig.\ \ref{numerical}, with the skyrmion
size shows the $\lambda_{\mathrm{eff}}^{2}$ trend with some logarithmic
contribution. Proportionality to $\lambda_{\mathrm{eff}}^{2}$ is
related to the fact that the magnetic moment of the skyrmion is roughly
proportional to the area inside which the spin field undergoes a significant
rotation. The logarithm comes from the magnetic moment of the tale
accumulated over a large area. Substitution of Eqs.\ (\ref{BP})
into the last term of Eq.\ (\ref{E-continuous}) gives for the Zeeman
energy of the skyrmion $E_{Z}=4\pi HS(\lambda/a)^{2}\ln(r_{{\rm max}}/\lambda)$,
where $r_{{\rm max}}\gg\lambda$ is the upper limit of integration
on $r$ determined by the size of the system, $L$, or the cutoff
due to the magnetic field, $\delta_{H}=a\sqrt{{JS}/{H}}$, whichever is smaller. In practice one always has $\delta_{H}\ll L$. 

For a skyrmion of size $\lambda\ll\delta_{H}$, a more rigorous approach
can be developed if one explicitly takes into account the screening
of the skyrmion profile by the magnetic field at $r\gg\lambda$, where
the problem can be linearized near $s_{z}=-1$. Splicing the asymptotic
solution for $\lambda\ll r$ with the BP solution, Eq. (\ref{BP}),
for $r\ll\delta_{H}$ and computing the integral over $r$ yields
\begin{eqnarray}
E_{Z} & = & 4\pi HS\left(\frac{\lambda}{a}\right)^{2}\left(\ln\frac{\delta_{H}}{\lambda}-\gamma+\ln2-\frac{1}{2}\right) \nonumber \\
& \simeq& 4\pi HS\left(\frac{\lambda}{a}\right)^{2}\ln\frac{0.681\delta_{H}}{\lambda},
\end{eqnarray}
where $\gamma=0.5772$ is the Euler constant. This formula requires
a strong inequality $\lambda\ll\delta_{H}$ that is difficult to
fulfill in practice. To extend the applicability range, one can add a constant
to the argument of the logarithm so that the resulting formula
fits most satisfactorily the Zeeman energy computed numerically
on the lattice, see Fig.\ \ref{numerical}. The best fit is provided by 
\begin{equation}
E_{Z}=4\pi HS\left(\frac{\lambda}{a}\right)^{2}\ln\left(1.5+0.68\frac{\delta_{H}}{\lambda}\right).\label{num-Zeeman} 
\end{equation}

In the continuous model we used for numerical work the value of log given by the above formula,
$l=\ln\left(1.5+0.68\delta_{H}/\lambda\right)$. At the critical (collapse) field, using Eq. (\ref{collapse}), one obtains $l=\ln\left[1.5+\left(1.06/l\right)\left(J/A\right)^{1/3}\right]$.
One can fit $H_{c}$ of Eq. (\ref{collapse}) to the numerical data
taking into account the dependence of the logarithm on $A$ and considering
1.5 and $1.06/l$ as fitting parameters. The best values of the latter
used in Fig. \ref{scaling} are 1.4 and 0.14.

\begin{thebibliography}{0}

\bibitem{SkyrmePRC58} 
T. H. R. Skyrme, Proc. Royal Soc. A \textbf{247}, 260-278 (1958).

\bibitem{BelPolJETP75} A. A. Belavin and A. M. Polyakov, Pis'ma Zh. Eksp. Teor. Fiz \textbf{22}, 503-506 (1975) {[}JETP Lett. \textbf{22},
245-248 (1975); A. M. Polyakov, \textit{Gauge Fields and Strings}, Harwood Academic Publishers 1987.

\bibitem{WiegmannPRL88} P. B. Wiegmann, Phys. Rev. Lett. \textbf{60},
821 (1988).

\bibitem{WenZeePRL88} X. G. Wen and A. Zee, Phys. Rev. Lett. \textbf{61},
1025 (1988).

\bibitem{HaldanePRL88} F. D. M. Haldane, Phys. Rev. Lett. \textbf{61},
1029 (1988).

\bibitem{ChaHalNelPRB89} S. Chakravarty, B. I. Halperin, and D. R.
Nelson, Phys. Rev. B \textbf{39}, 2344 (1989).

\bibitem{Lectures}
E. M. Chudnovsky and J. Tejada, {\it Lectures on Magnetism}, Rinton Press (Princeton - NJ, 2006).

\bibitem{Brown-book}
Book: {\it The Multifaceted Skyrmion}, edited by G. E. Brown and M. Rho (World Scientific, 2010).

\bibitem{AlkStoNat01} U. Al'Khawaja, and H. T. C. Stoof, Nature \textbf{411}, 918 (2001).

\bibitem{SonKarKivPRB93} S. L. Sondhi, A. Karlhede, S. A. Kivelson,
and E. H. Rezayi, Phys. Rev. B \textbf{47}, 16419 (1993).

\bibitem{StonePRB93} M. Stone, Phys. Rev. B \textbf{53}, 16573 (1996).

\bibitem{YeKimPRL99} Jinwu Ye, Y. B. Kim, A. J. Millis, B. I. Shraiman,
P. Majumdar, and Z. Tesanovic, Phys. Rev. Lett. \textbf{83}, 3737 (1999).

\bibitem{WriMerRMR89} D. C. Wright, and N. D. Mermin, Rev. Mod. Physi. \textbf{61}, 385 (1989).

\bibitem{Nagaosa2013}
N. Nagaosa and Y. Tokura, Nature Nanotech. {\bf 8}, 899 (2013).

\bibitem{Klaui2016}
G. Finocchio, F. B\"{u}ttner, R. Tomasello, M. Carpentieri, and M. Klaui, J. Phys. D: Appl. Phys. {\bf 49}, 423001 (2016).

\bibitem{Leonov-NJP2016}
A. O. Leonov, T. L. Monchesky, N. Romming, A. Kubetzka, A. N. Bogdanov, and R. Wiesendanger, New J. Phys. {\bf 18}, 065003 (2016).

\bibitem{Hoffmann-PhysRep2017}
W. Jiang, G. Chen, K. Liu, J. Zang, S. G. E. te Velthuis, and A. Hoffmann, Phys. Rep. {\bf 704}, 1 (2017). 

\bibitem{Fert-Nature2017}
A. Fert, N. Reyren, and V. Cros, Nature Rev. Mater. {\bf 2}, 17031 (2017).

\bibitem{Roldan-PRB2015}
A. Rold\`{a}n-Molina, M. J. Santander, A. S. Nunez, and J. Fern\`{a}ndez-Rossier, Phys. Rev. B {\bf 92}, 245436 (2015).

\bibitem{MQT-book}
Book: E. M. Chudnovsky and J. Tejada, {\it Macroscopic Quantum Tunneling of the Magnetic Moment} (Cambridge University Press, Cambridge - England, 1998). 

\bibitem{Springer}
Book: {\it Molecular Magnets, Physics and Applications}, edited by J. Bartolom\'{e}, Fernando Luis, and J. F. Fernandez (Springer, 2014).

\bibitem{Wernsdorfer}
M. Ganzhorn. S. Klyatskaya, M. Ruben, and W. Wernsdorfer, Nature Commun. {\bf 7}, 11443 (2016). 

\bibitem{Blatter}
G. Blatter, M. V. Feigel'man, V. B. Geshkenbein, A. I. Larkin, and V. M. Vinokur, Rev. Mod. Phys. {\bf 66}, 1125 (1994). 

\bibitem{Clarke}
J. Clarke and F. K. Wilhelm, Nature {\bf 453}, 1031(2008).

\bibitem{CCG-PRB2012}
L. Cai, E. M. Chudnovsky, and D. A. Garanin, Phys. Rev. B {\bf 86}, 024429 (2012).

\bibitem{IvanovPRB06} B. A. Ivanov, A. Y. Merkulov, V. A. Stepanovich,
C. E. Zaspel, Phys. Rev. B \textbf{74}, 224422 (2006).

\bibitem{IvanovPRB09} E. G. Galkina, E. V. Kirichenko, B. A. Ivanov,
V. A. Stephanovich, Phys. Rev. B \textbf{79}, 134439 (2009).

\bibitem{Moutafis-PRB2009}
C. Moutafis, S. Komineas, and J. A. C. Bland,  Phys. Rev. B {\bf 79}, 224429 (2009).

\bibitem{Ezawa-PRL2010}
M. Ezawa, Phys. Rev. Lett. {\bf 105}, 197202 (2010).

\bibitem{Makhfudz-PRL2012}
I. Makhfudz, B. Kr$\ddot{\text{u}}$ger, and O. Tchernyshyov, Phys. Rev. Lett. {\bf 109}, 217201 (2012).

\bibitem{AbanovPRB98} A. Abanov and V. L. Pokrovsky, Phys. Rev. B \textbf{58}, R8889 (1998).

\bibitem{Bogdanov-Nature2006}
U. K. R$\ddot{\text{o}}${\ss}ler, N. Bogdanov, and C. Pfleiderer,  Nature {\bf 442}, 797 (2006).

\bibitem{Heinze-Nature2011}
S. Heinze, 	K. von Bergmann, M. Menzel, J. Brede, A. Kubetzka, R. Wiesendanger, G. Bihlmayer, and S. Blugel, Nature Phys. {\bf 7}, 713 (2011).

\bibitem{Leonov-NatCom2015}
A. O. Leonov and M. Mostovoy, Nature Commun. {\bf 6}, 8275 (2015).

\bibitem{Lin-PRB2016}
S.-Z. Lin and S. Hayami, Phys. Rev. B {\bf 93}, 064430 (2016).

\bibitem{EC-DG-NJP2018}
E. M. Chudnovsky and D. A. Garanin, New J. Phys. {\bf 20}, 033006 (2018).

\bibitem{EC-PRA}
E. M. Chudnovsky, Phys. Rev. A {\bf 46}, 8011 (1992).

\bibitem{garchupro13}
D. A. Garanin, E. M. Chudnovsky, and T. C. Proctor, Phys. Rev. B {\bf 88}, 224418 (2013). 


\end{thebibliography}
\end{document}